\documentclass{iau}
\usepackage{graphicx}

\usepackage{natbib}
\usepackage{xspace}
\usepackage{hyperref}

\title[Modelling thermal X-rays around the Galactic centre] %% give here short title %%
{Modelling the thermal X-ray emission around the Galactic centre from \\colliding Wolf-Rayet winds}

\author[Christopher M.~P.~Russell, Q.~Daniel Wang \& Jorge Cuadra]   %% give here short author list %%
{Christopher M.~P.~Russell$^1$,
 Q.~Daniel Wang$^2$,
 \and Jorge Cuadra$^3$}

\affiliation{$^1$X-ray Astrophysics Laboratory, NASA/Goddard Space Flight Center,\\ Greenbelt, MD 20771, USA (NASA Postdoctoral Program Fellow, administered by USRA)\\ email: {\tt crussell@udel.edu} \\[\affilskip]
$^2$Department of Astronomy, University of Massachusetts, Amherst, MA 01003, USA% \\email: {\tt wqd@astro.umass.edu}
\\[\affilskip]
$^3$Instituto de Astrof\'{\i}sica, Facultad de F\'{\i}sica, Pontificia Universidad Cat\'{o}lica de Chile, 782-0436 Santiago, Chile%\\ email: {\tt jcuadra@astro.puc.cl}
}

\pubyear{2016}
\volume{322}  %% insert here IAU Symposium No.
\jname{The Multi-Messenger Astrophysics of the Galactic Centre}
\editors{S.~N.~Longmore, G.~Bicknell \& R.~Crocker, eds.}

\newcommand\arcsec{\hbox{$^{\prime\prime}$}}   % from mnras.cls
\newcommand\arcse{\hbox{$^{\prime\prime}$}\xspace}   % from mnras.cls, added \xspace portion myself
\newcommand*{\SAs}{{Sgr\,A$^*$}\xspace}

\begin{document}

\maketitle

\begin{abstract}
  The Galactic centre is a hotbed of astrophysical activity, with the injection of wind material from $\sim$30 massive Wolf-Rayet (WR) stars orbiting within 12\arcse of the super-massive black hole (SMBH) playing an important role.
  Hydrodynamic simulations of such colliding and accreting winds produce a complex density and temperature structure of cold wind material shocking with the ambient medium, creating a large reservoir of hot, X-ray-emitting gas.
  This work aims to confront the 3Ms of \textit{Chandra} X-ray Visionary Program (XVP) observations of this diffuse emission by computing the X-ray emission from these hydrodynamic simulations of the colliding WR winds, amid exploring a variety of SMBH feedback mechanisms.
  The major success of the model is that it reproduces the spectral shape from the 2\arcsec--5\arcse ring around the SMBH, where most of the stellar wind material that is ultimately captured by \SAs is shock-heated and thermalised.
  This naturally explains that the hot gas comes from colliding WR winds, and that the wind speeds of these stars are in general well constrained.
  The flux level of these spectra, as well as 12\arcsec$\times$12\arcse images of 4--9 keV, show the X-ray flux is tied to the SMBH feedback strength; stronger feedback clears out more hot gas, thereby decreasing the thermal X-ray emission.
  The model in which \SAs produced an intermediate-strength outflow during the last few centuries best matches the observations to within about 10\%, showing SMBH feedback is required to interpret the X-ray emission in this region.
  
\end{abstract}

\firstsection % if your document starts with a section,
\section{Introduction}
%%%%%%%%%%%%%%%%%%%%%%%%%%%%%%%%%%%%%%%%%%%%%%%%%%%%%%%%%%%%%%%%

The proximity of \SAs makes it the only SMBH where its orbiting stars are resolved, and therefore is the best opportunity to study the interplay between a SMBH and the stars and ejected wind material orbiting it.  Two examples of work in this thread are the hydrodynamic simulations of the winds of 30 Wolf-Rayet (WR) stars orbiting \SAs \citep{CuadraNayakshinMartins08,CuadraNayakshinWang15}, and the 3\,Ms of \textit{Chandra} X-ray Visionary Program (XVP) observations \citep{WangP13} that resolved, among other things, the diffuse thermal emission around \SAs thought to originate from wind-wind collisions.

This work computes the thermal X-ray emission from the aforementioned hydrodynamics simulations and compares the results to the XVP observations with the aim of increasing our understanding of the WR stars, and more generally the full environment, surrounding \SAs.  The full version of this work is \citet{RussellWangCuadra16}.

\newlength{\thisnum}
\setlength{\thisnum}{2.53cm}
\begin{figure*}
  \begin{flushleft}

  \includegraphics[height=\thisnum]{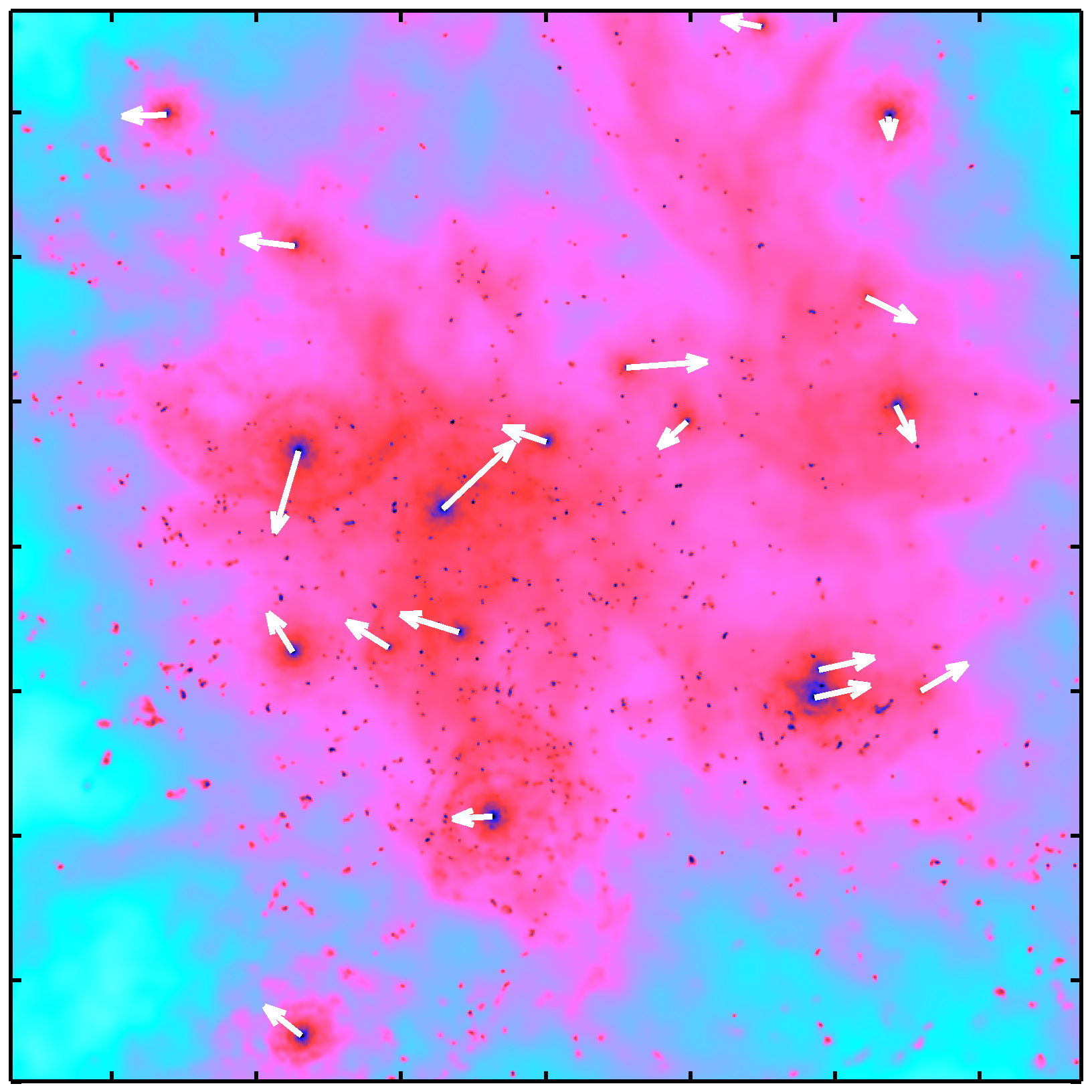}
  \put(-70,7){\scriptsize \fontfamily{phv}\selectfont \textbf{NF}}
  \put(-30,18){\scriptsize \fontfamily{phv}\selectfont \textbf{IRS~13E}}
  \includegraphics[height=\thisnum]{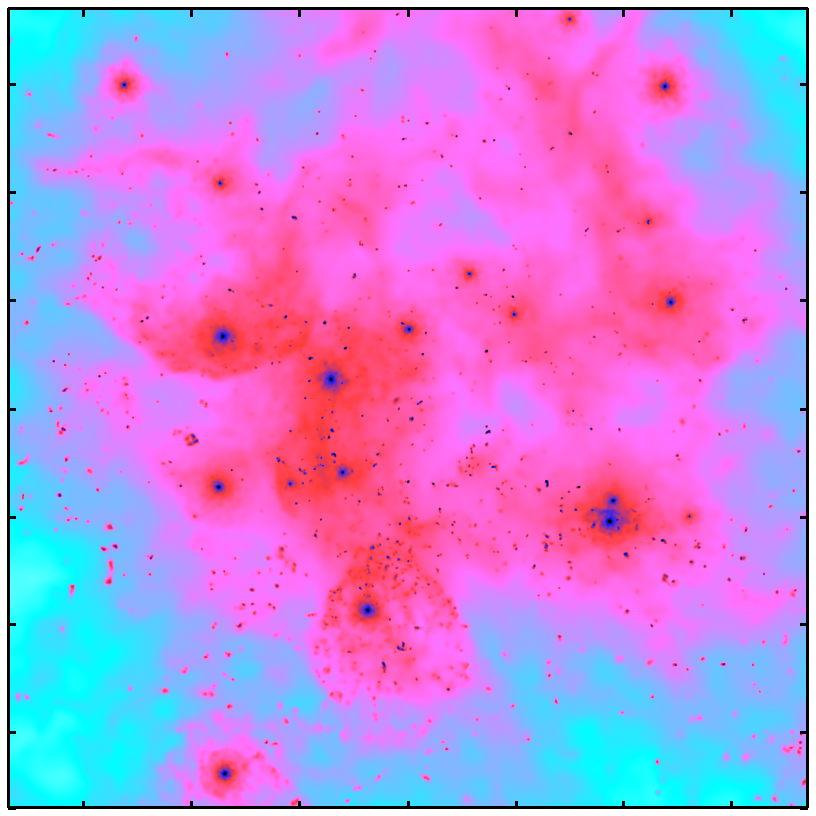}\put(-70,7){\scriptsize \fontfamily{phv}\selectfont \textbf{OF}}%
  \includegraphics[height=\thisnum]{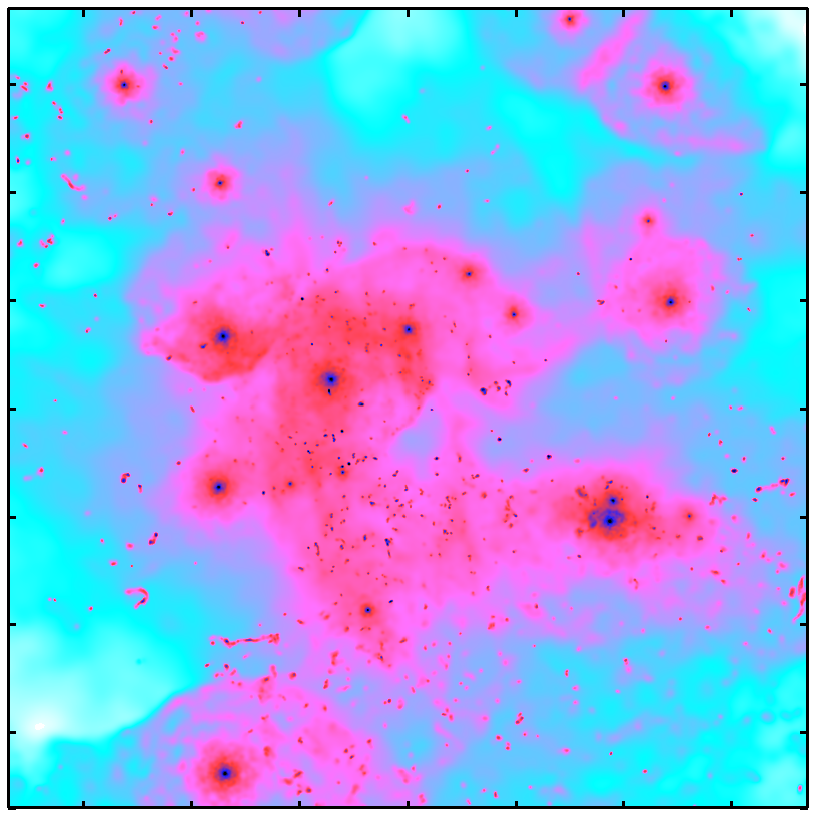}\put(-70,7){\scriptsize \fontfamily{phv}\selectfont \textbf{OBBP}}
  \put(-42,66){\vector(-1,0){29.5}}
  \put(-30,66){\vector(1,0){29.5}}
  \put(-41.5,64){\scriptsize \fontfamily{phv}\selectfont \textbf{12\arcsec}}
  \includegraphics[height=\thisnum]{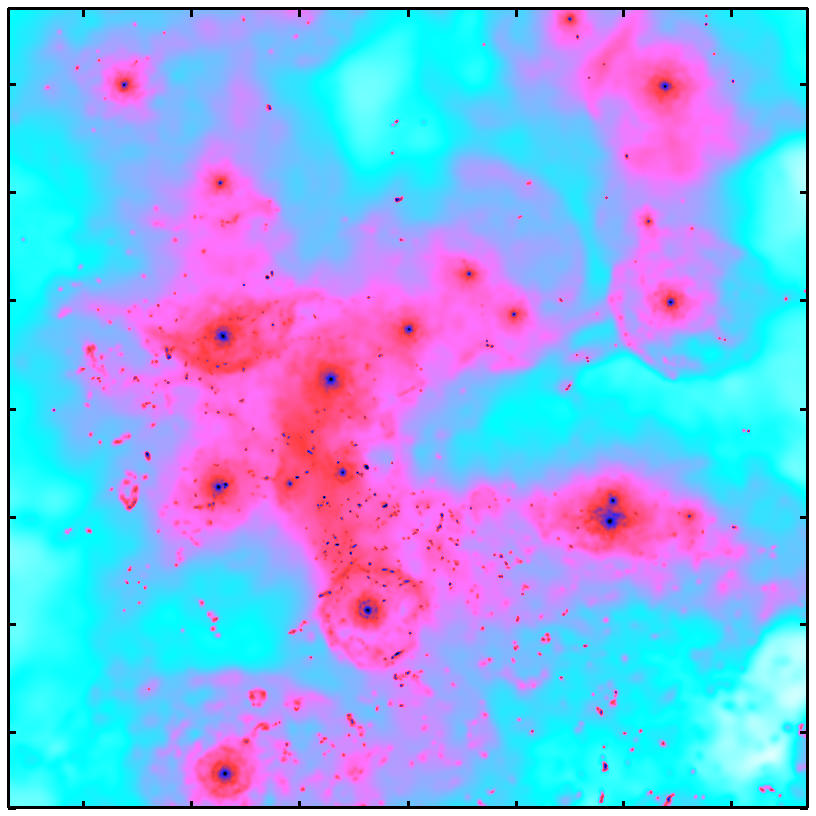}\put(-70,7){\scriptsize \fontfamily{phv}\selectfont \textbf{OB5}}%
  \includegraphics[height=\thisnum]{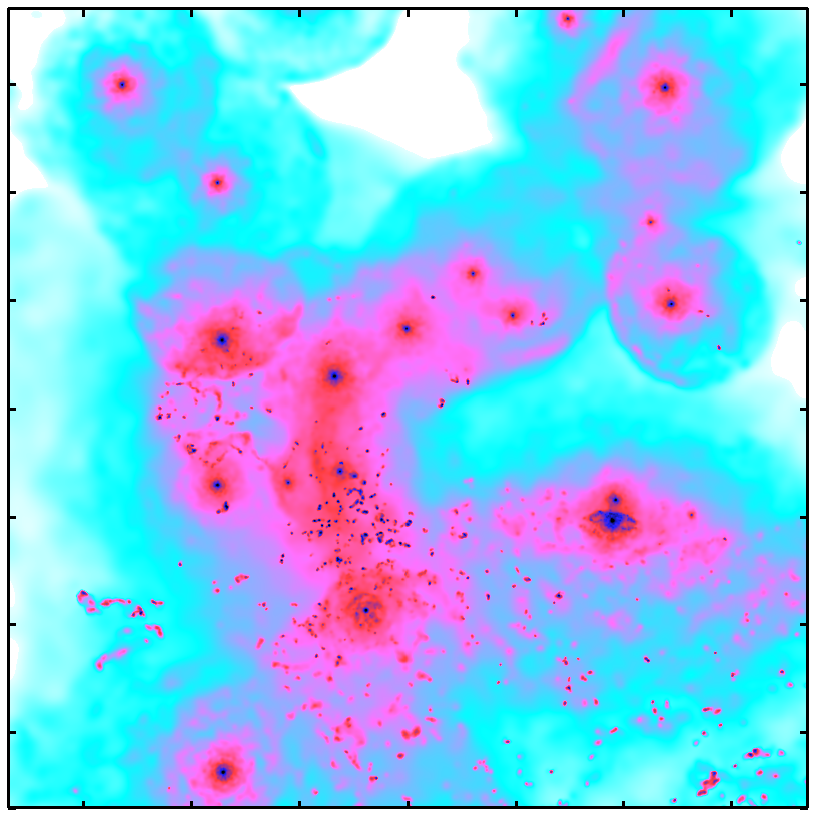}\put(-70,7){\scriptsize \fontfamily{phv}\selectfont \textbf{OB10}}%
  \includegraphics[height=\thisnum]{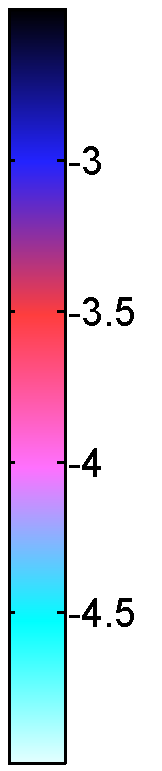}\put(0,22){\scriptsize \fontfamily{phv}\selectfont \rotatebox{90}{log g/cm\textsuperscript{\tiny 2}}}

  \caption{Column density of the central 12\arcsec$\times$12\arcse for all models with the strength of the SMBH feedback increasing from left to right.  The arrows indicate the projected velocities of the stars.  The tick marks are every 2$\times$10$^{17}$ cm, and the IRS~13E cluster is labeled.}
  \label{fi:SPHcoldens}
  \end{flushleft}
\end{figure*}
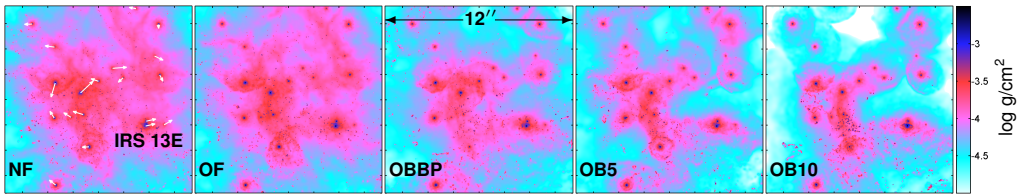
\begin{table}
  \centering
  \caption{\textit{Left:} Parameters of SMBH feedback for each model.  \textit{Center:} Results of the ISM absorption fitting over the entire spectrum of the 2\arcsec--5\arcse ring (excluding IRS13E and the PWN), showing the inverse norm (i.e.\ [model flux]/[observed flux]), $N_{\rm H}$ with 90\% confidence interval errors, and $\chi^2_{\rm red}$ for each model.  Each fit has 211 \textit{dof}. \textit{Right:} Model-to-data ratio of the 4-9 keV flux from 2\arcsec--5\arcse ring (excluding IRS13E and the PWN).  The differences in the image and spectral results account for the different X-ray background estimations for each observable, resulting in the observed image flux being an upper limit and the observed spectral flux being a lower limit.}
  \label{ta}
  \vspace{0.15cm}
  %{\scriptsize
  \begin{tabular}{l | c c c | c c c | c c c}
    model & $\dot{M}_{\rm out}$ & $v_{\rm out}$ & direction           & norm$^{-1}$ & $N_{\rm H}$             & $\chi^2_{\rm red}$ & image & spectra & mean \\
          & ($M_\odot$/yr)      & (km/s)        &                     & (mod/obs)   & (10$^{22}$\,cm$^{-2}$)  &                    & \multicolumn{3}{c}{(model/observed)}      \\ \hline
    NF    & 0                   & 0             & --                  & 1.95        & 10.28$^{+0.29}_{-0.28}$ & 1.36               & 1.50  & 1.97    & 1.73 \\
    OF    & $\dot{M}_{\rm in}$  & 10,000        & spherical           & 1.94        & 10.26$^{+0.29}_{-0.28}$ & 1.36               & 1.50  & 1.96    & 1.72 \\
    OBBP  & $10^{-4}$           & 5,000         & bipolar, 15$^\circ$ & 1.20        & 10.16$^{+0.29}_{-0.27}$ & 1.38               & 0.92  & 1.21    & 1.07 \\
    OB5   & $10^{-4}$           & 5,000         & spherical           & 1.17        & 10.92$^{+0.30}_{-0.28}$ & 1.31               & 0.92  & 1.20    & 1.06 \\
    OB10  & $10^{-4}$           & 10,000        & spherical           & 0.79        & 10.73$^{+0.31}_{-0.29}$ & 1.24               & 0.62  & 0.81    & 0.71 \\
  \end{tabular}%}
\end{table}

%%%%%%%%%%%%%%%%%%%%%%%%%%%%%%%%%%%%%%%%%%%%%%%%%%%%%%%%%%%%%%%%
\section{Method}\label{M}
%%%%%%%%%%%%%%%%%%%%%%%%%%%%%%%%%%%%%%%%%%%%%%%%%%%%%%%%%%%%%%%%

Fig.~\ref{fi:SPHcoldens} shows the column density at the present day for all hydrodynamic simulations \citep{CuadraNayakshinMartins08,CuadraNayakshinWang15}.  As the feedback strength increases from left to right, the amount of material remaining in the simulation volume decreases.
The model names are NF -- no feedback, OF -- outflow, OBBP -- bipolar outburst, OB5 -- 5,000 km/s outburst, and OB10 -- 10,000 km/s outburst.  All outbursts occur from 400 to 100 yr ago \citep{PontiP10}. Table \ref{ta} (left) provides more details of each model.

We solve the formal solution of radiative transfer along a 500$\times$500 grid of rays covering the central 15\arcsec$\times$15\arcse through the simulation volume for 0.3-12 keV (covering the \textit{Chandra} HETG response function) at a resolution of 800 energy bins per dex.  These pixel maps are folded through the \textit{Chandra} ACIS-S/HETG 0th-order response function to compare with the observed spectrum, and then through the \textit{Chandra} PSF (which is approximated as a 0.5\arcse full-width half-max [FWHM] Gaussian) to compare with the observed image.

The basis of the radiative transfer calculation (see \citealt{Russell13}; \citealt{RussellP16} for more details) is the SPH visualization program \textsc{splash} \citep{Price07}.  The X-ray emissivities are from the \texttt{VVAPEC} model \citep{SmithP01} using \texttt{AtomDB} version 2.0.2, as implemented in \texttt{XSpec} \citep{Arnaud96} version 12.0.9c.  The wind opacities are from
\citet{VernerYakovlev95} (obtained using the interface of \citealt{LeuteneggerP10}), and the interstellar medium (ISM) opacities are from the \texttt{TBabs} model \citep{WilmsAllenMcCray00}.  As the emissivities and wind opacities are metallicity dependent, we use three models to cover the range of WR spectral types in the SPH simulations; one for the WC stars \citep{Crowther07}, one for WN5-7 \citep{OniferP08}, and one for WN8-9 and Ofpe/WN9 (\texttt{CMFGEN} website).  Given that the high ISM column density obscures $E<1$\,keV, the optical depths over the observable range are sufficiently low such that the radiative transfer is well into the optically thin limit.

The only free parameter in the model is the ISM absorbing column, $N_{\rm H}$.  Table \ref{ta} (center) shows the fitting results for each model obtained with \texttt{XSpec}.  The best-fit value is $N_{\rm H}$\,=\,1.1$\times$10$^{23}$\,cm$^{-2}$, which we use to make all X-ray images and spectra in this paper.

%%%%%%%%%%%%%%%%%%%%%%%%%%%%%%%%%%%%%%%%%%%%%%%%%%%%%%%%%%%%%%%%
\section{Results}\label{R}
%%%%%%%%%%%%%%%%%%%%%%%%%%%%%%%%%%%%%%%%%%%%%%%%%%%%%%%%%%%%%%%%

Figs.~\ref{fi:imFB}~\&~\ref{fi:spFB} (left) show the 4--9 keV images and the 2\arcsec--5\arcse ring spectra for all models.  To compare the model and data images more quantitatively, Fig.~\ref{fi:spFB} (right) shows the 4--9 keV intensity as a function of projected radius from \SAs.

\setlength{\thisnum}{2.56cm}
\begin{figure}
  \begin{flushleft}

  \includegraphics[height=\thisnum]{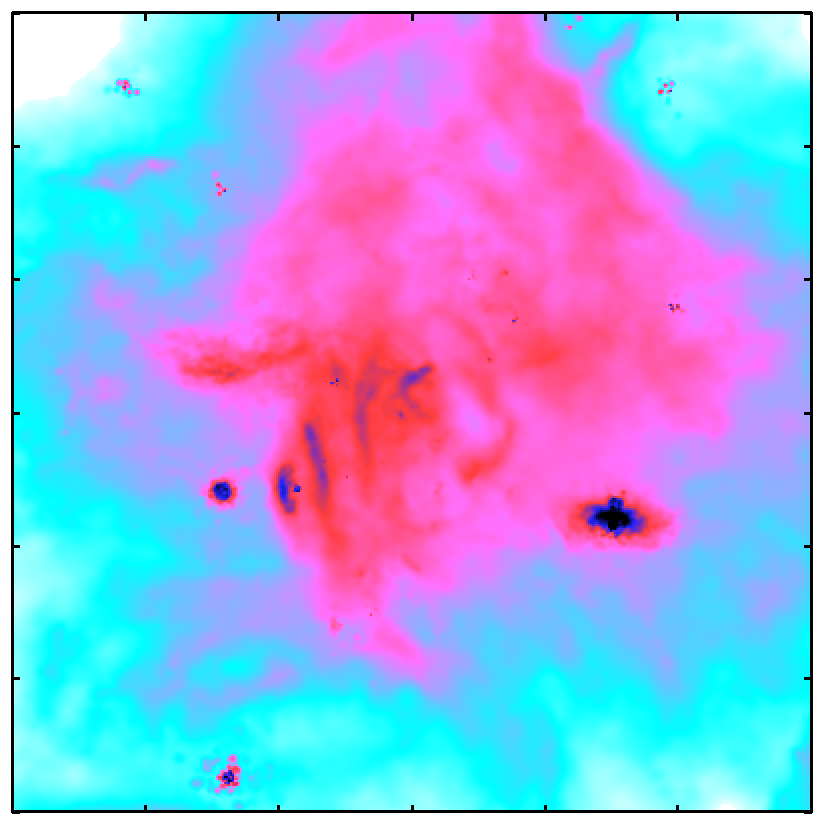}
  \put(-13,10){\scriptsize \fontfamily{phv}\selectfont \textbf{OF}}
  \put(-27.7,3){\scriptsize \fontfamily{phv}\selectfont \textbf{no PSF}}%
  \includegraphics[height=\thisnum]{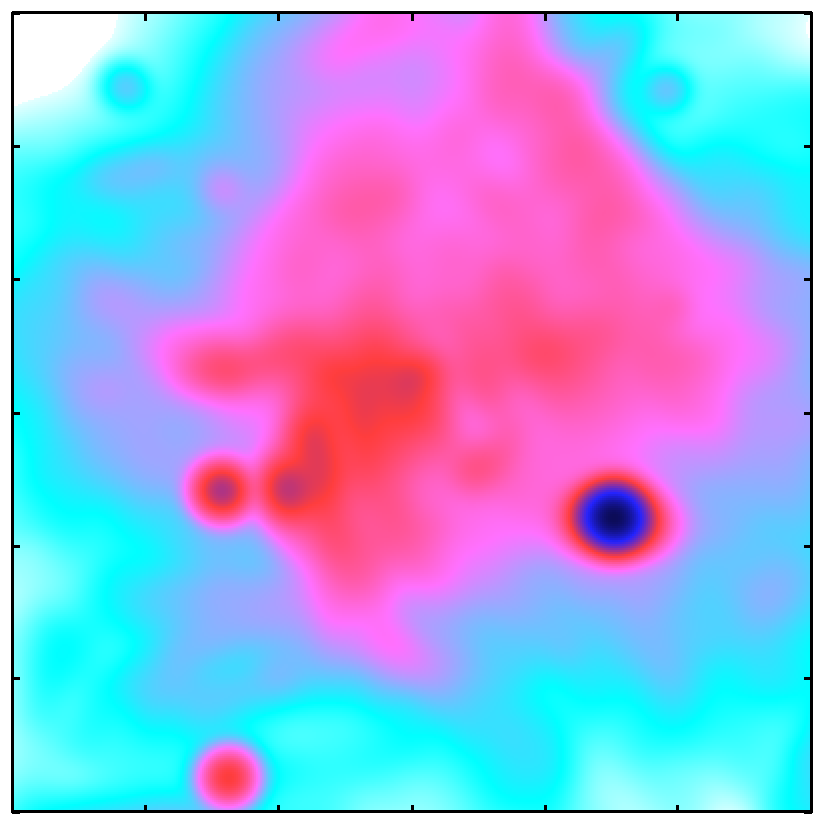}
  \put(-13,10){\scriptsize \fontfamily{phv}\selectfont \textbf{OF}}
  \put(-32.4,3){\scriptsize \fontfamily{phv}\selectfont \textbf{with PSF}}%
  \includegraphics[height=\thisnum]{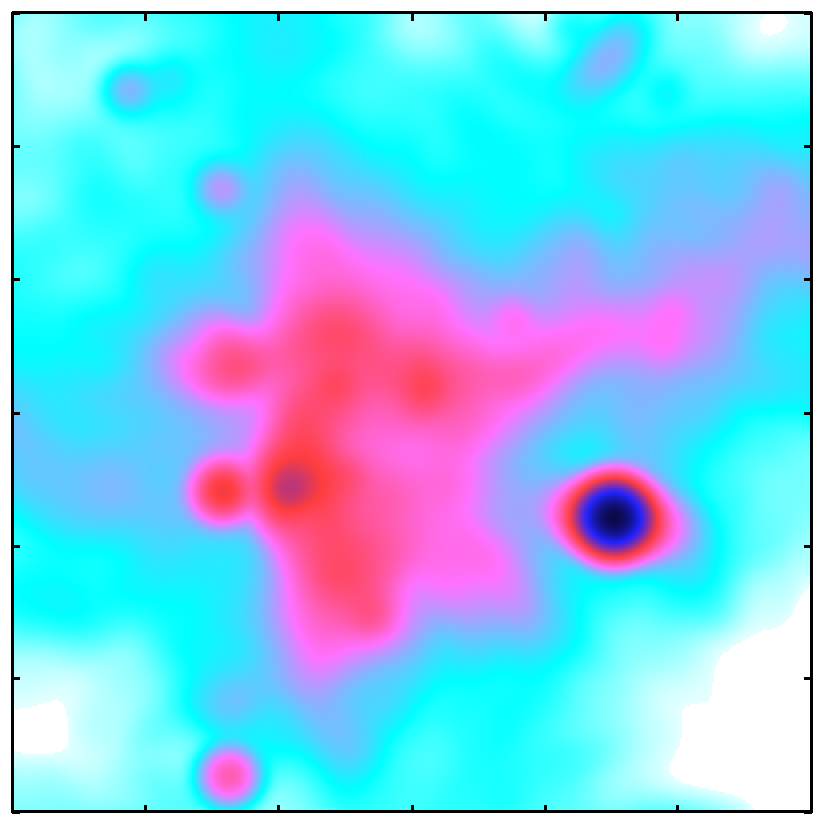}
  \put(-22,10){\scriptsize \fontfamily{phv}\selectfont \textbf{OBBP}}
  \put(-32.4,3){\scriptsize \fontfamily{phv}\selectfont \textbf{with PSF}}
  \put(-42,67){\vector(-1,0){30}}
  \put(-31,67){\vector(1,0){30}}
  \put(-42,65){\scriptsize \fontfamily{phv}\selectfont \textbf{12\arcsec}}%
  \includegraphics[height=\thisnum]{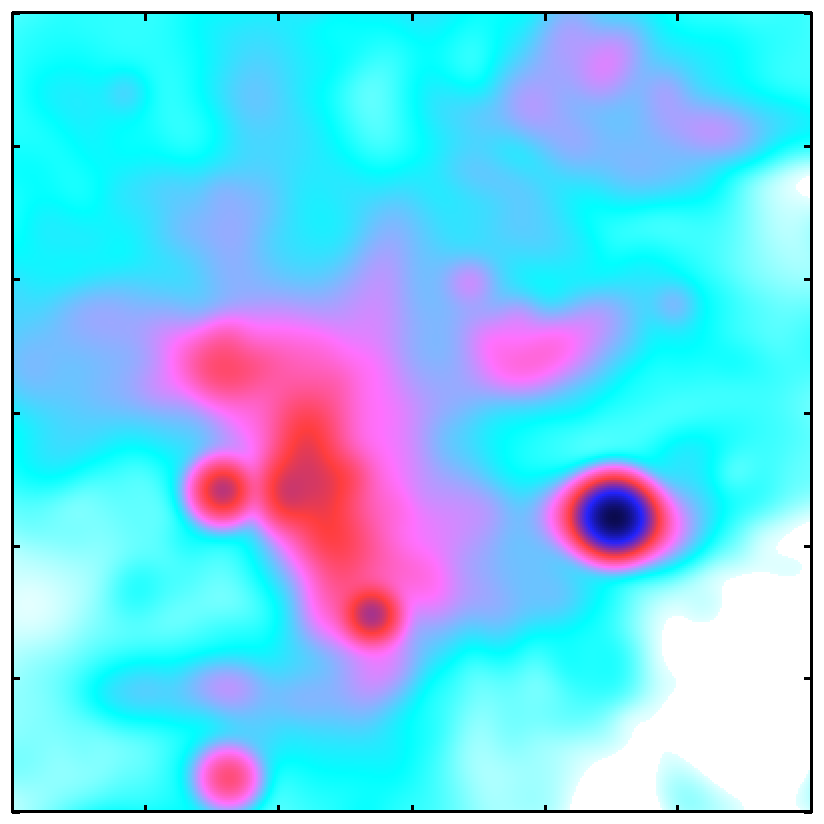}
  \put(-17.4,10){\scriptsize \fontfamily{phv}\selectfont \textbf{OB5}}
  \put(-32.4,3){\scriptsize \fontfamily{phv}\selectfont \textbf{with PSF}}%
  \includegraphics[height=\thisnum]{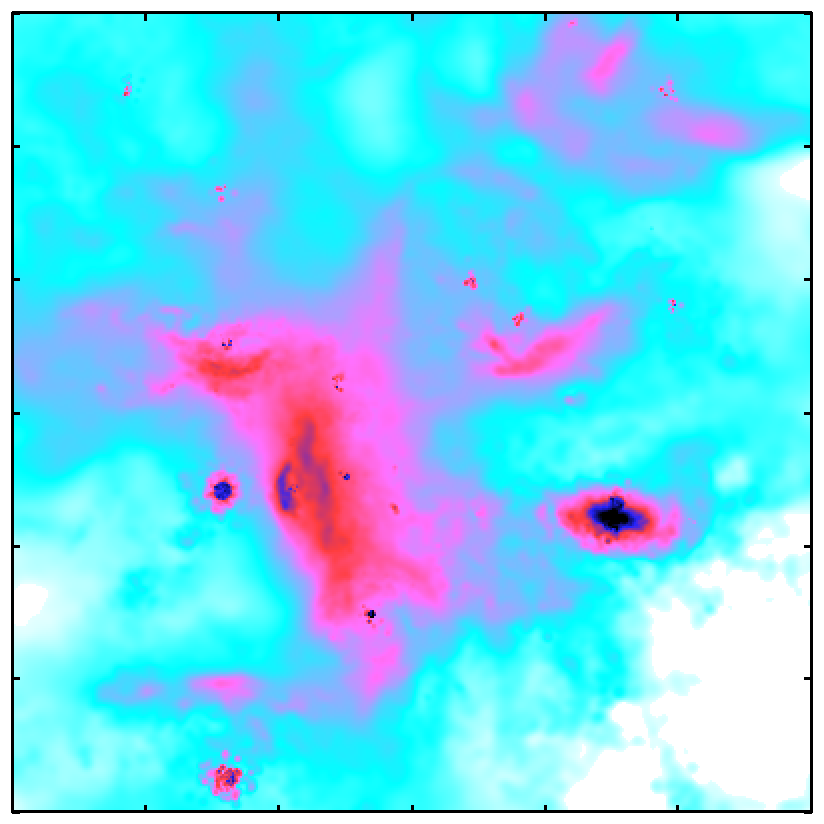}
  \put(-17.4,10){\scriptsize \fontfamily{phv}\selectfont \textbf{OB5}}
  \put(-27.7,3){\scriptsize \fontfamily{phv}\selectfont \textbf{no PSF}}%
  \includegraphics[height=\thisnum]{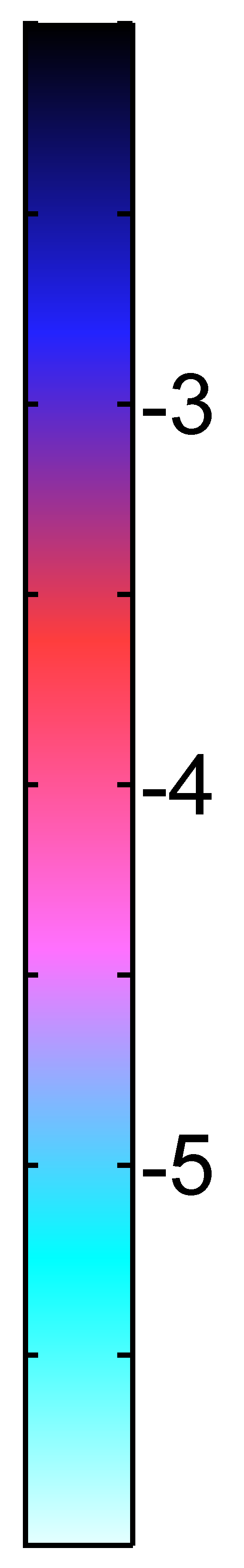}
  \put(0,11){\scriptsize \fontfamily{phv}\selectfont \rotatebox{90}{log cnt/s/arcsec\textsuperscript{\tiny 2}}}

  \vspace{-0.04cm}

  \includegraphics[height=\thisnum]{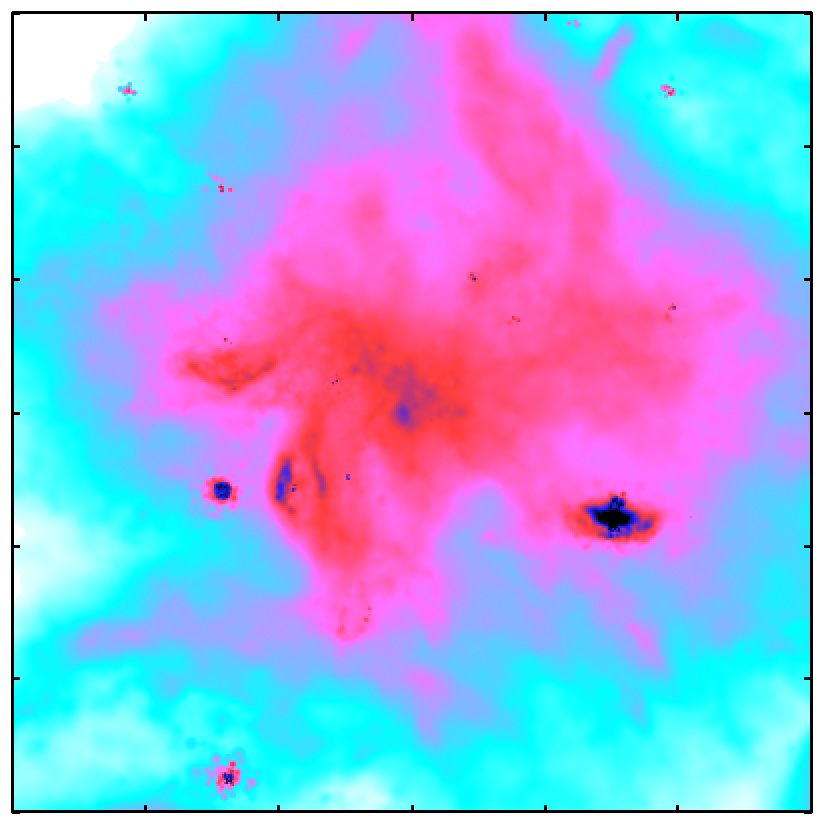}
  \put(-13.4,10){\scriptsize \fontfamily{phv}\selectfont \textbf{NF}}
  \put(-27.7,3){\scriptsize \fontfamily{phv}\selectfont \textbf{no PSF}}
  \put(-32,20){\scriptsize \fontfamily{phv}\selectfont \textbf{IRS~13E}}%
  \includegraphics[height=\thisnum]{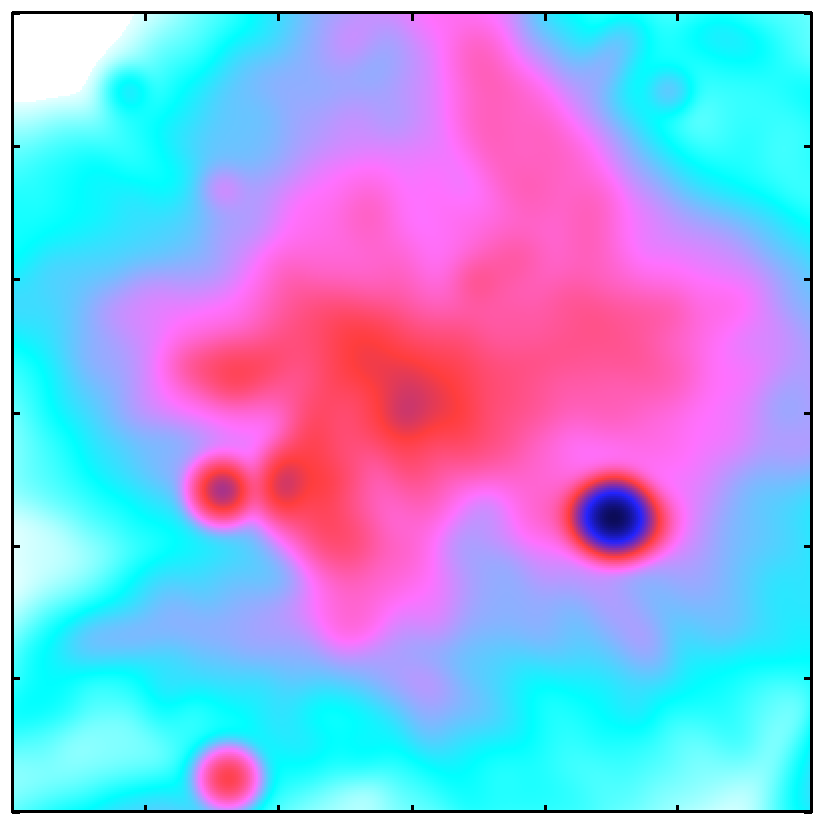}
  \put(-13.4,10){\scriptsize \fontfamily{phv}\selectfont \textbf{NF}}
  \put(-32.4,3){\scriptsize \fontfamily{phv}\selectfont \textbf{with PSF}}%
  \includegraphics[height=\thisnum]{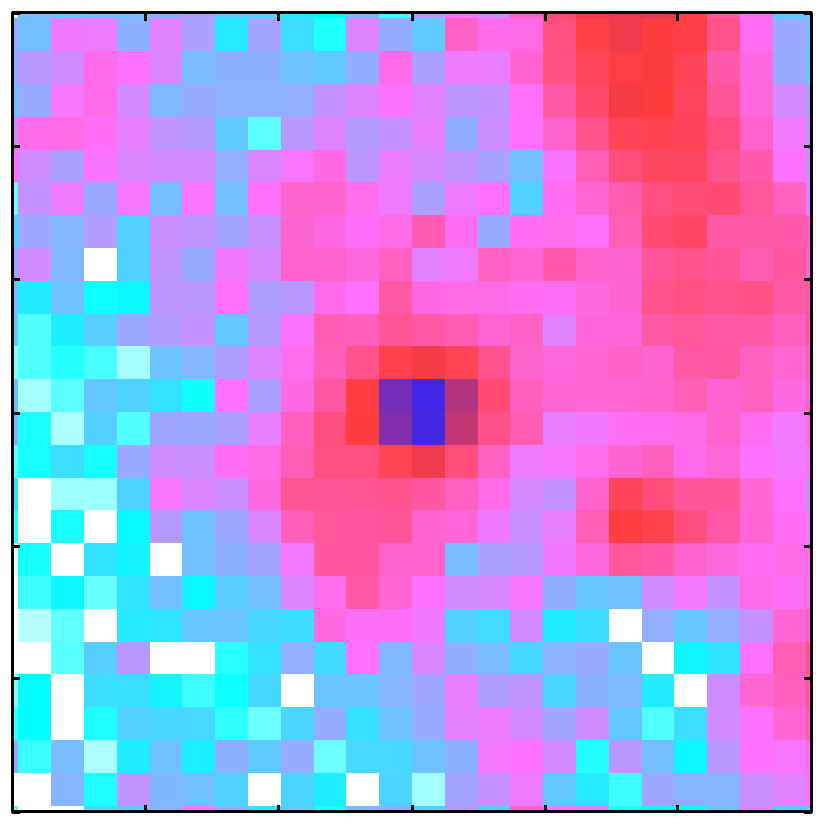}
  \put(-17.6,3){\scriptsize \fontfamily{phv}\selectfont \textbf{data}}
  \put(-22,67){\scriptsize \fontfamily{phv}\selectfont \rotatebox{300}{\textbf{PWN}}}%
  \includegraphics[height=\thisnum]{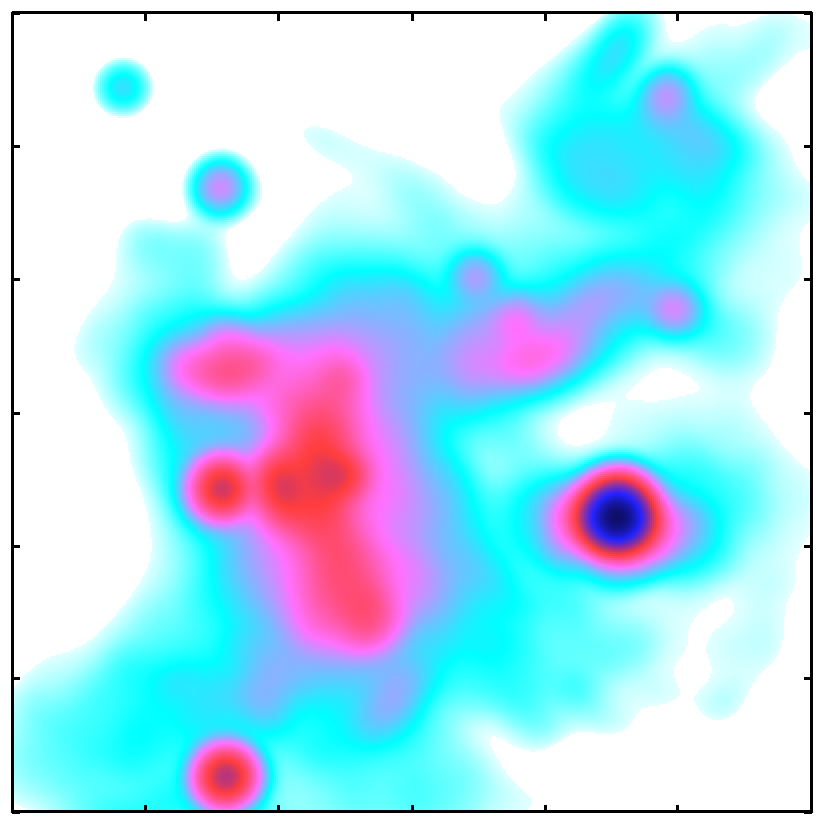}
  \put(-20.7,10){\scriptsize \fontfamily{phv}\selectfont \textbf{OB10}}
  \put(-32.4,3){\scriptsize \fontfamily{phv}\selectfont \textbf{with PSF}}%
  \includegraphics[height=\thisnum]{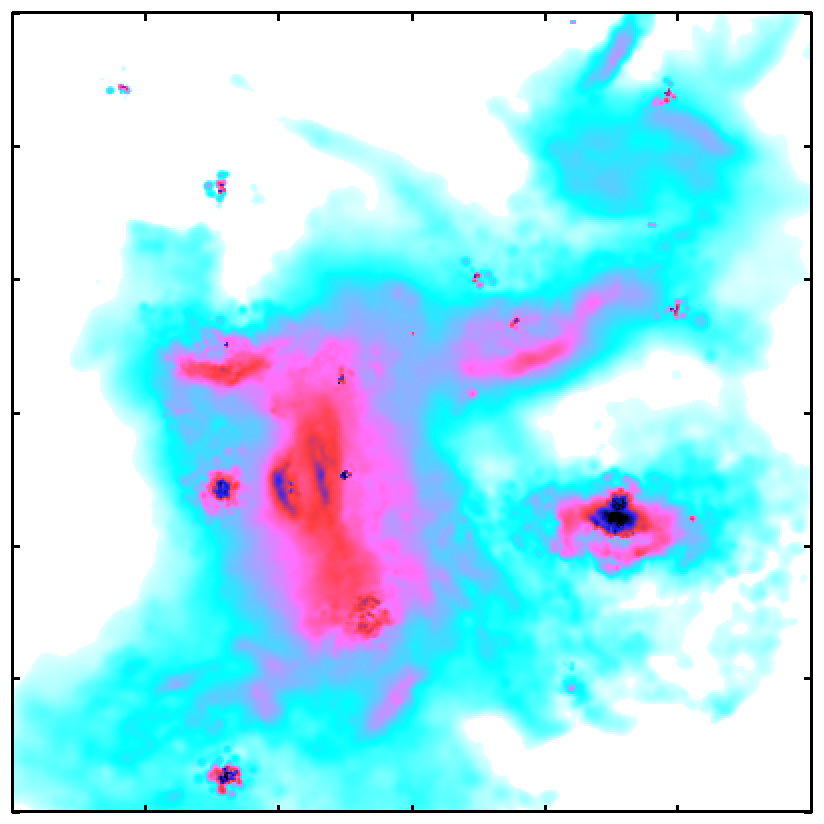}
  \put(-20.7,10){\scriptsize \fontfamily{phv}\selectfont \textbf{OB10}}
  \put(-27.7,3){\scriptsize \fontfamily{phv}\selectfont \textbf{no PSF}}%
  \includegraphics[height=\thisnum]{GCImage_ColorBar_4to9keV_BigFont_NSTC.png}
  \put(0,11){\scriptsize \fontfamily{phv}\selectfont \rotatebox{90}{log cnt/s/arcsec\textsuperscript{\tiny 2}}}
  \caption{\textit{Chandra} 4--9 keV ACIS-S/HETG 0th-order images (12\arcsec$\times$12\arcsec) comparing all models with the observation, which is in the bottom centre panel.  For the models, the first and last columns are not folded through the PSF, while the 3 central columns are.  The SMBH feedback strength increases clockwise (NF to OB10), while the X-ray flux decreases, showing that the clearing out of material by the SMBH feedback (Fig.~\ref{fi:SPHcoldens}) affects the thermal X-ray emission.}
  \label{fi:imFB}
  \end{flushleft}
\end{figure}

\begin{figure}
  \includegraphics[width=16pc]{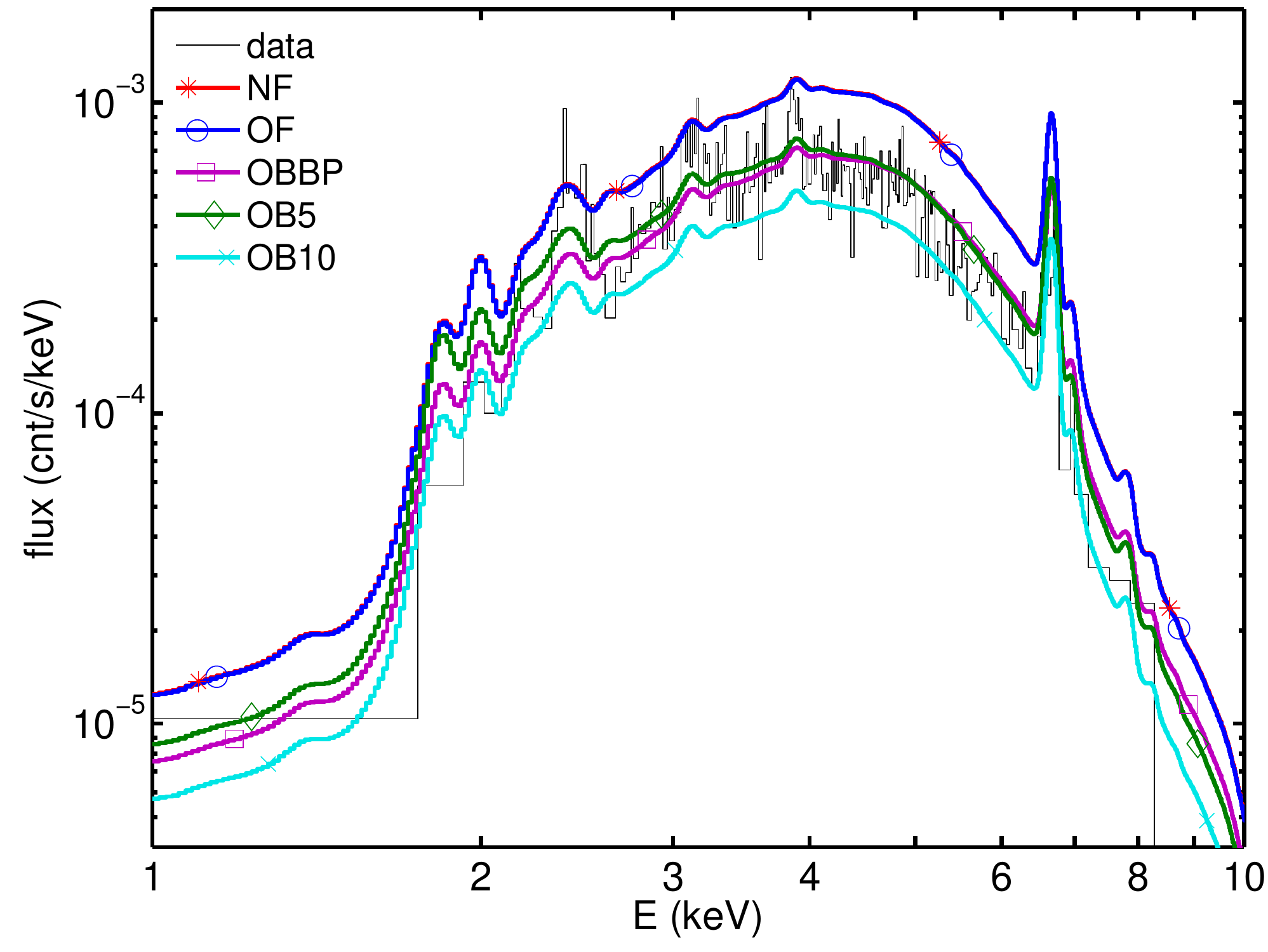}%
  \includegraphics[width=16pc]{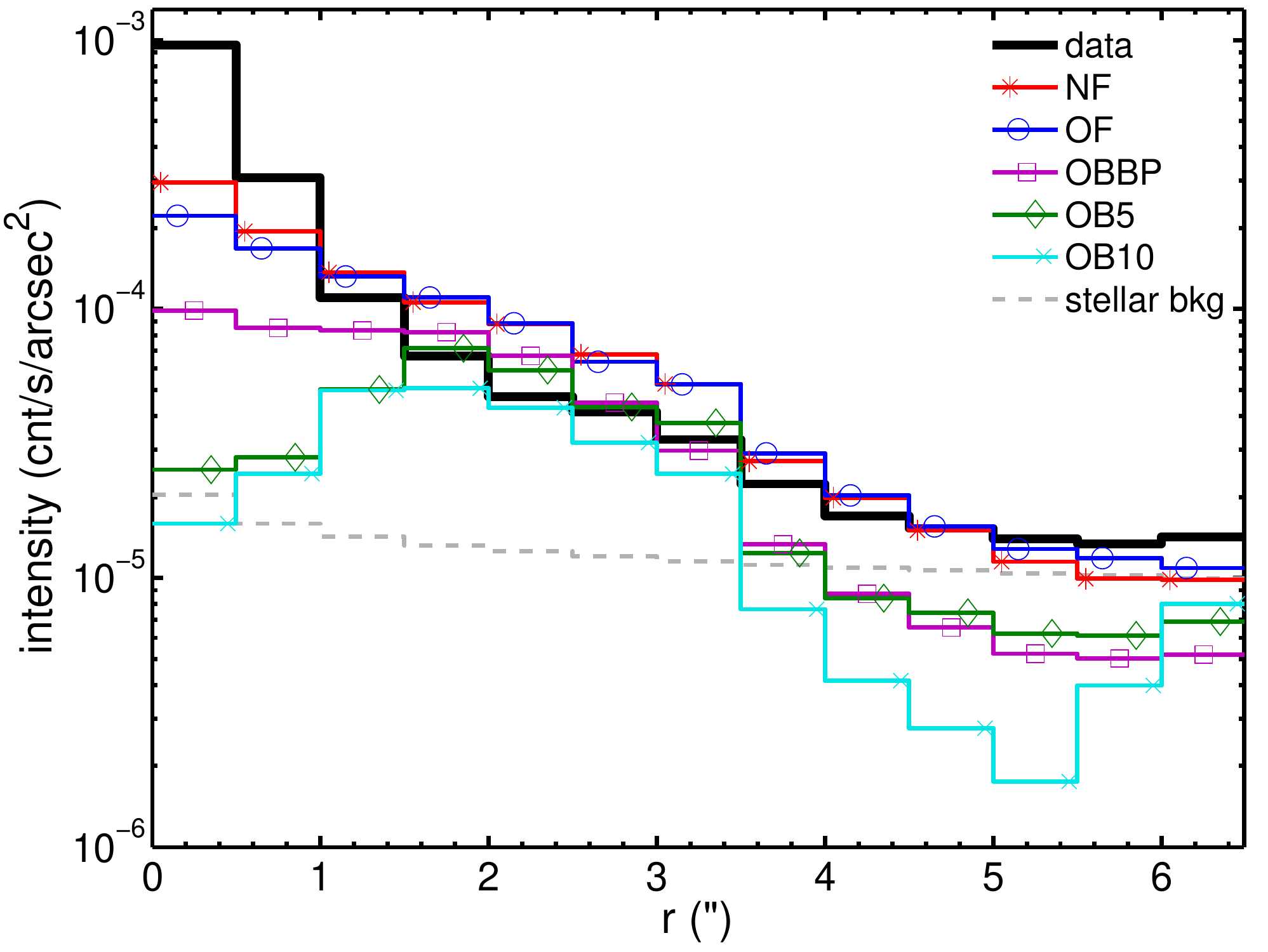}
  \caption{\textit{Left:} \textit{Chandra} ACIS-S/HETG 0th-order spectra of the models and observation from the 2\arcsec--5\arcse ring centred on \SAs (excluding the IRS~13E and PWN contributions).  As expected, the level of X-ray flux follows the trend of Fig~\ref{fi:imFB}. \textit{Right:} 4--9 keV intensity of the observation and models as a function of radius from \SAs.  The background component due to the stellar population of CVs, which has been subtracted from the observational image, is shown.}
  \label{fi:spFB}
\end{figure}

A noteworthy achievement of the model is the agreement in the spectral shape (Fig.~\ref{fi:spFB}, left).  This provides strong evidence that the WR wind-wind collisions are the dominant source of X-ray emission around \SAs.  The temperature of the gas around \SAs is naturally explained by the WR shocked material, validating the wind speeds of the WR stars used in the simulations.

The model that best agrees with the observation over 2\arcsec--5\arcse (excluding IRS~13E and the PWN) is the medium-strength feedback model OB5 (see Table \ref{ta}, right).  The ratio of the model-to-data flux over 4--9~keV is 0.92 and 1.20 in the image and spectral comparison, yielding a mean discrepancy of the model flux being $\sim$6\% higher than the data.

The largest disagreement is the IRS~13E cluster (model/data $\sim$\,7), indicating that its WC and WN winds should be revised downward to agree with the X-ray observations.

%%%%%%%%%%%%%%%%%%%%%%%%%%%%%%%%%%%%%%%%%%%%%%%%%%%%%%%%%%%%%%%%
\section{Conclusions}\label{C}
%%%%%%%%%%%%%%%%%%%%%%%%%%%%%%%%%%%%%%%%%%%%%%%%%%%%%%%%%%%%%%%%

We compute the thermal X-ray properties of the Galactic centre from hydrodynamic simulations of the 30 WR stars orbiting  within 12\arcse of \SAs \citep{CuadraNayakshinMartins08,CuadraNayakshinWang15}.  These simulations use different feedback models from the SMBH at its centre.  The \textit{Chandra} X-ray Visionary Program observations \citep{WangP13} provide an anchor point for these simulations, so we compare the observed 12\arcsec$\times$12\arcse 4--9 keV image and the 2\arcsec--5\arcse ring spectrum with the same observables synthesized from the models.  Remarkably, the shape of the model spectra, regardless of the type of feedback, agree very well with the data.  This indicates the hot gas around \SAs is primarily from shocked WR wind material, and that the velocities of these winds are well constrained.  The X-ray flux strongly depends on the feedback mechanism; greater SMBH outflows clear out more WR-ejected material around \SAs, thus decreasing the model X-ray emission.  Over 4--9~keV in energy and 2\arcsec--5\arcse in projected distance from \SAs (excluding IRS~13E and the nearby PWN), the X-ray emission from all models is within a factor of 2 of the observations, with the best model agreeing to within 10\%; this is the medium-strength feedback model OB5, which has an SMBH outburst of $\dot{M}_{\rm out}=10^{-4}$\,$M_\odot$\,yr$^{-1}$ and $v$\,=\,5000\,km\,s$^{-1}$ from 400 to 100\,yr ago. Therefore, this work shows that the SMBH outburst is required for fitting the X-ray data, and by extension that the outburst still affects the current X-ray emission around \SAs, even though it ended 100\,yr ago.

%%%%%%%%%%%%%%%%%%%%%%%%%%%%%%%%%%%%%%%%%%%%%%%%%%%%%%%%%%%%%%%%%

%%%%%%%%%%%%%%%%%%%%%%%%%%%%%%%%%%%%%%%%%%%%%%%%%%%%%%%%%%%%%%%%%

\end{document}